\documentclass{article}

\usepackage[final]{neurips_2021}

\usepackage[utf8]{inputenc} 
\usepackage[T1]{fontenc}    
\usepackage{hyperref}       
\usepackage{url}            
\usepackage{booktabs}       
\usepackage{amsfonts}       
\usepackage{nicefrac}       
\usepackage{microtype}      
\usepackage{xcolor}         
\usepackage{natbib}
\usepackage{graphicx}
\title{Mapping Access to Water and Sanitation in Colombia using Publicly Accessible Satellite Imagery, Crowd-sourced Geospatial Information and Random Forests}

\author{%
  Niccolo Dejito, Ren Avell Flores, Rodolfo de Guzman, Isabelle Tingzon \\
  Thinking Machines Data Science\\
  Taguig City, Philippines \\
  \texttt{\{cholo, avell, ram, issa\}@thinkingmachin.es} \\
  \And
  Liliana Carvajal, Alberto Aroca, Carlos Delgado \\
  IMMAP Colombia\\
  Bogota, Colombia \\
  \texttt{\{lcarvajal, aaroca, cdelgado\}@immap.org} \\
}

\begin{document}

\maketitle

\begin{abstract}
    Up-to-date, granular, and reliable quality of life data is crucial for humanitarian organizations to develop targeted interventions for vulnerable communities, especially in times of crisis. One such quality of life data is access to water, sanitation and hygeine (WASH). Traditionally, data collection is done through door-to-door surveys sampled over large areas. Unfortunately, the huge costs associated with collecting these data deter more frequent and large-coverage surveys. To address this challenge, we present a scalable and inexpensive end-to-end WASH estimation workflow using a combination of machine learning and census data, publicly available satellite images, and crowd-sourced geospatial information. We generate a map of WASH estimates at a granularity of 250m x 250m across the entire country of Colombia. The model was able to explain up to 65\% of the variation in predicting access to water supply, sewage, and toilets. The code is made available with MIT License at https://github.com/thinkingmachines/geoai-immap-wash.
    
\end{abstract}

\section{Application Context}
The context of this paper is the Venezuelan migration crisis, wherein Venezuelans informally settle in neighboring Colombia to escape a devastating economic crisis that decimated the value of their earnings to the point where basic human needs are rendered unaffordable even with a monthly income. With 5.6 million migrants from a population of 30 million, the crisis has turned into one of the largest mass migrations in the region throughout its history. Despite the direness of the situation however, aid for migrants remains underfunded. Basic human needs such as access to water, sanitation, and hygiene (WASH) remain unaddressed.
Solving WASH access is an important and pressing concern because prolonged exposure to unsanitary and unhygienic practices can lead to disease. To address these needs, humanitarian organizations should know where WASH access is lowest. These statistics are reported by the government, but not frequently enough to quickly respond to the ongoing crisis.

Together with our partner iMMAP Colombia, a non-profit organization working closely with the local government, we locate settlements that need most help by generating granular maps of WASH access across Colombia using satellite images, crowdsourced geodata, and machine learning. Our solution is impactful because it presents a method to infer areas that need water facilities in the absence of reliable data, especially in previously unpopulated locations \citep{tingzon2020mapping} that are now inhabited out of need by displaced persons. With a clear basis for impact, we can help direct limited funds more effectively and encourage donation with transparency. Concepts that may be of interest to the general ML practitioner include (1) the WASH metrics, which are used in epidemiology studies and policy decisions to measure quality of life; (2) census in developing countries, which are infrequent due to high costs to do them; and (3) freely available data such as satellite-captured images like nighttime lights and labelled locations from OpenStreetMap, which have been shown to provide information value in model-based inference. \citep{ledesma2020interpretable}

\section{Introduction}
Access to water, sanitation, and hygiene (WASH) is a key measure to quality of life. It is defined by proximity to and availability of infrastructures in place to support basic human needs such as having potable water, proper sewerage, and proper toilets. Lack of access to basic water and sanitation facilities has been linked to illnesses \citep{joseph_2019}, stunted child growth \citep{sanitation2015sanitation}, and hampered economic development \citep{world2017wake}. For these reasons, increasing access to WASH is listed as Goal 6 of the United Nations Sustainable Development Goals.

However, the need for granular and timely data is crucial for actionable decisions on humanitarian relief and aid operations. Though included in government reports, WASH access data are often not granular enough to delineate vulnerable areas from prosperous regions and not timely enough to respond to current events. These reports are usually sampled on select locations within a large area, conducted infrequently, and aggregated at the municipal level. These data limitations stem from the large costs and extensive manpower required to conduct such surveys.

In Colombia, the need for granular and timely data has been emphasized further by the Venezuelan migration crisis. Since 2014, Colombia has been experiencing an influx of refugees from neighboring Venezuela, in response to worsening living conditions brought about by hyperinflation and corruption. These refugees take shelter in informal settlements near the outskirts of urban areas, areas that are not registered and included by the urban-focused census. However, the WASH census takes place every 13 years, and the next one will be too late to respond to the needs of those displaced by the crisis.

To fill in data gaps for aggregated and dated census, cost-efficient methods from more readily available data sources have been an active topic of research. Previous works have achieved success predicting wealth statistics from satellite imagery using computer vision techniques \citep{burke2020using}. Other works also investigated methods for estimating epidemiology measures for tracking down risk factors of illnesses such as malaria  \citep{gething2011new} and cholera \citep{joseph_2019}. From these papers, a pioneering work \citep{andres2018geo} by the World Bank applied a geostatistical model on Nigeria to measure WASH access. Adapting ideas from this paper, we estimate WASH for the whole of Colombia using available census data and freely available satellite images. 

The main contributions of this paper are as follows:
\begin{itemize}
\item We generate a WASH access map for Colombia at a resolution of 250m x 250m, a much higher resolution compared to how WASH census data are normally made available;
\item We show the effectiveness of using publicly available satellite imagery, crowd-sourced geospatial information, and random forests for modeling WASH access;
\item We develop a scalable and inexpensive end-to-end workflow for estimating WASH using census data and freely available satellite image data, the code for which is made available with MIT License at https://github.com/thinkingmachines/geoai-immap-wash.
\end{itemize}

\section{Related Work}
For mapping WASH access, data collection is often done through on-the-ground household surveys. Tools such as mobile or computer applications facilitate the centralization of household information for later aggregation. For example, in WASH monitoring engagements directed towards small communities in Zimbabwe \citep{ngala_2017} and India \citep{batchelor2013using}, researchers coordinated with survey teams to collect data on water points and catchment areas using handheld GPS devices. The survey team also interviewed the local community on their day-to-day experience accessing these infrastructures, such as the length of time it takes to reach the facilities,  the frequency of which water points break down in response to seasonal rainfall, and the size of the household, in the form of questionnaires. In the latest survey of WASH tools \citep{schweitzer2014mapping}, the majority of WASH tools were developed to serve specific communities and lack an easily replicable format, taking weeks to deploy in other areas.  It is clear from these projects that survey data collection is a labor-intensive, unscalable, and expensive task.

To address these challenges, recent works have turned to a class of models called Model-Based Geostatistics (“MBG”). MBG is a generalized linear mixed model that considers spatial autocorrelation (which relates values to values that are located nearby) to produce more accurate predictions. The studies that use this method come from various contexts like epidemiological ones such as malaria risk mapping \citep{gething2011new}, and cholera outbreak prevention \citep{joseph_2019}, which use satellite images as geospatial covariates to predict the prevalence in populations and display the most afflicted on a map. A recent work by Deshpande, et al. \citep{deshpande2020mapping} also demonstrates a wide-scale multi-national study applying MBG to produce WASH access maps on a municipal level, providing benchmarks for 88 low to middle-income countries globally. Methods using traditional machine learning (ML) methods are not as widely used, perhaps owing to the fact that these bodies of literature have a rich history while ML has only become widespread in recent years.

A similar study was done in the World Bank WASH estimation project in Nigeria \citep{andres2018geo}. In the study, authors developed an MBG model to generate WASH access maps for the whole of Nigeria, at a resolution of 1kmx1km. The target variables were taken from the Nigerian National Water and Sanitation Survey which collected access data on households nationwide as well as geo-located water points and water schemes. The predictor variables used include freely accessible satellite images as follows: vegetation, aridity, temperature, nighttime lights, travel time, and separately, population for calculating the amount of affected. They evaluated the performance using a cross-validation procedure of random 75-25 train-test split evaluated over 4 folds.

\section{Data}
In this study, we used the WASH census data as our primary source of ground truth against which model prediction will be evaluated. For features, we used satellite images and Points of Interest (POIs). We standardize all data by aligning and resampling to 250m x 250m grids using QGIS. 

\subsection{WASH Census as Labels}
We use the percentage of households with no access to (1) water supply, (2) sewage, and (3) toilets as training labels. These percentages were derived from 2018 census data aggregated to a block or district level, provided by the Colombian government’s statistics office. The data includes WASH information on both urban and rural households. The urban data is aggregated on a block level, collected via door-to-door surveys. The rural data consists of data calculated from surveyed water points and the community’s access to these specific water points. We used these data points to get the finest available granularity, but note that these differences in aggregation and collection methods may create slight differences in value ranges for urban versus rural areas.

\subsection{Satellite Images as Features}
We download the following satellite imagery products for the year 2018 via the Google Earth Engine \citep{gorelick2017google} platform: vegetation, aridity, temperature, nighttime lights, population, elevation, urban change. Similar to previous studies \citep{andres2018geo}, we select images based on how well they explain urbanization, development, and distance to water sources. For more details on each image such as explanation, resolution, and publishing agencies, please see the Appendix.

\subsection{Points of Interest as Features}
We extract the following POI types from OpenStreetMap, via GeoFabrik in August 2020: waterways (natural, i.e. river, stream, or manmade, i.e. canals, ditches, dams), commercial (e.g. offices, retail stores), restaurants, hospitals, airports, highways (including primary, trunk and motorway).

\section{Methods}

\begin{figure}[h!]
\centering
\includegraphics[width=0.5\textwidth]{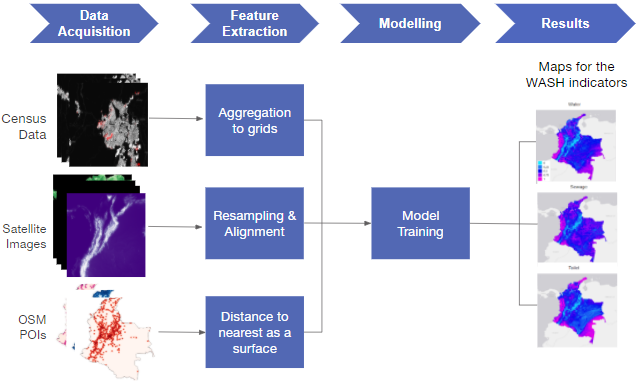}
\caption{An end-to-end workflow for estimating WASH.}
\end{figure}

\subsection{Feature Extraction}
We prepare the dataset for training by performing the following preprocessing steps.

\begin{enumerate}
\item Aggregate census to grid - we acquire 57,036 grid cells for training the model by transforming the dataset to grid level, using the following steps: (1) calculate the centroid of each block; (2) map centroids to grid-based on intersection; (3) recalculate percentage of households per grid based on block membership; and (4) filter out grid blocks whose surface area comprise mostly of blocks with no data (more than 50 percent).

\item Satellite image resampling and alignment - Preprocessing includes median aggregation across the whole period of 1 year for 2018, resampling to a standardized 250m x 250m grid, and normalizing values to (0, 1).

\item Conversion of POI to distance surfaces - Since the POIs come in the form of longitude, latitude pairs, we generate a raster map by calculating the distance to the nearest POI type for each of the centroids of the 250m x 250m grid cells. A particular location, denoted by the tile centroid, is considered to have low access to water if its distance to the nearest water source is greater than 5 km.  All spatial operations were done using Google BigQuery.
\end{enumerate}

\subsection{Model Training and Evaluation}
We use a Random Forest Regressor and evaluate using a cross-validation procedure of random 80-20 train-test split evaluated over 5 folds. Random forests are faster with large-scale datasets, considering the lack of neighbor-to-neighbor calculation, which is an intermediate step in geostatistical model fitting. The method performed on par or better using the same validation method these papers used. Specifically, the training of the model took 3 minutes in a 4CPU 15GB RAM computer.

We validate our results numerically by comparing accuracy metrics versus the benchmark paper \citep{andres2018geo}. We select the three metrics from the 7 WASH indicators (1) Piped water on premises, which checks if water is available directly on household premises via pipes; (2) Sewerage connection which checks if a household is connected to a sewer system; and (3) Open defecation which checks if a household has no access to any type of toilet.

\section{Results and Discussion}
We predict percentage of households with no WASH access. The previous study \citep{andres2018geo} achieved scores within the range of 60-70\% explained variation (r-squared) for WASH estimation in Nigeria. We find our results to perform comparably, with metrics shown in Table \ref{table:distribution}.

\begin{table}[h]
\caption{Validation Statistics}
\begin{center}
\begin{tabular}{lcc}
\hline
\multicolumn{1}{l}{\textbf{Variable}} & \multicolumn{1}{c}{\textbf{\begin{tabular}[c]{@{}c@{}}R-Squared\end{tabular}}} &
\multicolumn{1}{c}{\textbf{\begin{tabular}[c]{@{}c@{}}Root Mean\\ Squared Error\end{tabular}}} \\
\hline
Access to water & 0.5967 & 0.1291 \\ 
Access to sewage & 0.6474 & 0.1707 \\ 
Access to toilets & 0.5357 & 0.1233 \\ 
\hline
\end{tabular}
\label{table:distribution}
\end{center}
\end{table}

We note the following sources of errors: (1) different grid sizes, (2) difference in urban and rural data collection, (3) model error. As noted previously, differences in aggregation and collection methods for our ground truth data limit model performance. First, the census data available for the study is aggregated on a block-level, government-defined boundaries which do not perfectly match the 250m x 250m grids used for modelling. Second, the rural data was calculated from surveyed water points, which is in contrast with the urban data which used door-to-door surveys. Rural areas consist of large spaces with sparse housing, and as such estimates for these areas may be misrepresented by point samples. For our study, we included both urban and rural data for training as this setup yielded the highest accuracy from our experiments. 

Improving the model can be done by (1) increasing the time period covered by the dataset, (2) aligning the granularity of labels to features, and (3) including other relevant geospatial covariates. First, adding more datasets from a different year allows the model to account for temporal variability. In this study we are limited to the use of 2018 data because only 2018 has available block-level government data. Second, aggregating household data to the 250m x 250m grid size aligns the data to the features and ensures comparability across areas, by skipping the centroid intersection step in Section Feature Extraction. Understandably, information on households are aggregated to anonymize survey participants. By working closely with the government and sending them the grid polygons to aggregate on, it will be possible to aggregate these locations while still preserving anonymity. Third, including other relevant geospatial covariates may also improve model results. As satellite imagery become more available over time, better datasets may provide new ways to explain the variation.

We note the following ethical considerations for our work. The resulting areas tagged to have low WASH access can potentially be used to discriminate, degrade, or embarrass people living in them. The granular 250m grid cell can potentially identify specific households in that cell, considering the spread of settlements in sparse rural areas. To safeguard against these cases of misuse, we restrict access to both training data and model results, to be made available only after requesting from our partner humanitarian organization. We hope that justification towards an independent, benevolent third party allows room for careful consideration, prevents misuse, and ensures that our work is used for its intended purpose: to help those in need.

\section{Conclusion}
In summary, we have demonstrated the effectiveness of using a random forest machine learning model to estimate WASH access. From the resulting model we generate a high-resolution map measuring access to water, sewage and toilets in Colombia, at a resolution of 250m x 250m. Finally, we make the source code openly available.

The model will be used by iMMAP Colombia, a humanitarian organization that provides data relevant for key decision-making for on-the-ground efforts such as delivering aid and relief to the most vulnerable communities in Colombia. Using this model, they can locate areas which have low WASH access for the whole of Colombia, and prioritize accordingly. We emphasize that this method seeks not to replace the ground surveys but to fill the time gaps that these surveys are not always able to provide. Future works can improve model usability by increasing the period covered by the data set, aligning the granularity of labels to features, and including other relevant geospatial covariates.

\section*{Acknowledgment}
This project was done as a collaborative effort between Thinking Machines, and iMMAP Colombia, with the financing of the Office of U.S. Foreign Disaster Assistance (OFDA) of USAID. We acknowledge the support of Pia Faustino and Ardie Orden and thank them for the insightful discussions.

\bibliography{mybibliography}

\begin{thebibliography}{15}
\providecommand{\natexlab}[1]{#1}
\providecommand{\url}[1]{\texttt{#1}}
\expandafter\ifx\csname urlstyle\endcsname\relax
  \providecommand{\doi}[1]{doi: #1}\else
  \providecommand{\doi}{doi: \begingroup \urlstyle{rm}\Url}\fi

\bibitem[Andres et~al.(2018)Andres, Bhatt, Dasgupta, Echenique, Gething,
  Grabinsky~Zabludovsky, and Joseph]{andres2018geo}
Luis~A Andres, Samir Bhatt, Basab Dasgupta, Juan~A Echenique, Peter~W Gething,
  Jonathan Grabinsky~Zabludovsky, and George Joseph.
\newblock \emph{Geo-spatial modeling of access to water and sanitation in
  Nigeria}.
\newblock The World Bank, 2018.

\bibitem[Andrés et~al.(2017)Andrés, Duret, Mantovani, Molini, and
  Ort]{world2017wake}
Luis Andrés, Michel Duret, Pier Mantovani, Vasco Molini, and Rachel Ort.
\newblock \emph{A Wake Up Call: Nigeria Water Supply, Sanitation, and Hygiene
  Poverty Diagnostic}.
\newblock 06 2017.

\bibitem[Batchelor(2013)]{batchelor2013using}
James Batchelor.
\newblock Using gis and swat analysis to assess water scarcity and wash
  services levels in rural andhra pradesh, 2013.

\bibitem[Burke et~al.(2020)Burke, Driscoll, Lobell, and Ermon]{burke2020using}
Marshall Burke, Anne Driscoll, David Lobell, and Stefano Ermon.
\newblock Using satellite imagery to understand and promote sustainable
  development.
\newblock Technical report, National Bureau of Economic Research, 2020.

\bibitem[Deshpande et~al.(2020)Deshpande, Miller-Petrie, Lindstedt, Baumann,
  Johnson, Blacker, Abbastabar, Abd-Allah, Abdelalim, Abdollahpour,
  et~al.]{deshpande2020mapping}
Aniruddha Deshpande, Molly~K Miller-Petrie, Paulina~A Lindstedt, Mathew~M
  Baumann, Kimberly~B Johnson, Brigette~F Blacker, Hedayat Abbastabar, Foad
  Abd-Allah, Ahmed Abdelalim, Ibrahim Abdollahpour, et~al.
\newblock Mapping geographical inequalities in access to drinking water and
  sanitation facilities in low-income and middle-income countries, 2000--17.
\newblock \emph{The Lancet Global Health}, 8\penalty0 (9):\penalty0
  e1162--e1185, 2020.

\bibitem[Gething et~al.(2011)Gething, Patil, Smith, Guerra, Elyazar, Johnston,
  Tatem, and Hay]{gething2011new}
Peter~W Gething, Anand~P Patil, David~L Smith, Carlos~A Guerra, Iqbal~RF
  Elyazar, Geoffrey~L Johnston, Andrew~J Tatem, and Simon~I Hay.
\newblock A new world malaria map: Plasmodium falciparum endemicity in 2010.
\newblock \emph{Malaria journal}, 10\penalty0 (1):\penalty0 378, 2011.

\bibitem[Gorelick et~al.(2017)Gorelick, Hancher, Dixon, Ilyushchenko, Thau, and
  Moore]{gorelick2017google}
Noel Gorelick, Matt Hancher, Mike Dixon, Simon Ilyushchenko, David Thau, and
  Rebecca Moore.
\newblock Google earth engine: Planetary-scale geospatial analysis for
  everyone.
\newblock \emph{Remote sensing of Environment}, 202:\penalty0 18--27, 2017.

\bibitem[Joseph(2019)]{joseph_2019}
George Joseph.
\newblock Cholera data and modelling in planning \& targeting, 2019.
\newblock URL
  \url{https://www.fondation-merieux.org/wp-content/uploads/2018/10/6th-annual-meeting-of-gtfcc-george-joseph.pdf}.

\bibitem[Ledesma et~al.(2020)Ledesma, Garonita, Flores, Tingzon, and
  Dalisay]{ledesma2020interpretable}
Chiara Ledesma, Oshean~Lee Garonita, Lorenzo~Jaime Flores, Isabelle Tingzon,
  and Danielle Dalisay.
\newblock Interpretable poverty mapping using social media data, satellite
  images, and geospatial information.
\newblock \emph{arXiv preprint arXiv:2011.13563}, 2020.

\bibitem[Lundberg and Lee(2017)]{lundberg2017unified}
Scott~M Lundberg and Su-In Lee.
\newblock A unified approach to interpreting model predictions.
\newblock In \emph{Advances in neural information processing systems}, pages
  4765--4774, 2017.

\bibitem[Ngala(2017)]{ngala_2017}
Parvin Ngala.
\newblock Geospatial mapping of wash infrastructure and services, May 2017.
\newblock URL
  \url{https://www.emma-toolkit.org/report/geospatial-mapping-wash-infrastructure-and-services}.

\bibitem[Schweitzer et~al.(2014)Schweitzer, Grayson, and
  Lockwood]{schweitzer2014mapping}
Ryan Schweitzer, Claire Grayson, and Harold Lockwood.
\newblock Mapping of water, sanitation and hygiene sustainability tools.
\newblock \emph{Working Paper 10, IRC/Aguaconsult/Triple-S}, 2014.

\bibitem[Team et~al.(2015)Team, Team, Humphrey, Jones, Manges, Mangwadu,
  Maluccio, Mbuya, Moulton, Ntozini, et~al.]{sanitation2015sanitation}
Sanitation Hygiene Infant Nutrition Efficacy (SHINE)~Trial Team, Sanitation
  Hygiene Infant Nutrition Efficacy (SHINE)~Trial Team, Jean~H Humphrey,
  Andrew~D Jones, Amee Manges, Goldberg Mangwadu, John~A Maluccio, Mduduzi~NN
  Mbuya, Lawrence~H Moulton, Robert Ntozini, et~al.
\newblock The sanitation hygiene infant nutrition efficacy (shine) trial:
  rationale, design, and methods.
\newblock \emph{Clinical Infectious Diseases}, 61\penalty0 (suppl\_7):\penalty0
  S685--S702, 2015.

\bibitem[Tingzon et~al.(2020)Tingzon, Dejito, Flores, Guzman, Carvajal, Erazo,
  Cala, Villaveces, Rubio, and Ghani]{tingzon2020mapping}
Isabelle Tingzon, Niccolo Dejito, Ren~Avell Flores, Rodolfo~De Guzman, Liliana
  Carvajal, Katerine~Zapata Erazo, Ivan Enrique~Contreras Cala, Jeffrey
  Villaveces, Daniela Rubio, and Rayid Ghani.
\newblock Mapping new informal settlements using machine learning and time
  series satellite images: An application in the venezuelan migration crisis.
\newblock \emph{CoRR}, abs/2008.13583, 2020.
\newblock URL \url{https://arxiv.org/abs/2008.13583}.

\bibitem[Trabucco and Zomer(2019)]{trabucco_zomer_2019}
Antonio Trabucco and Robert Zomer.
\newblock Global aridity index and potential evapotranspiration (et0) climate
  database v2, Jan 2019.
\newblock URL
  \url{https://figshare.com/articles/dataset/Global_Aridity_Index_and_Potential_Evapotranspiration_ET0_Climate_Database_v2/7504448/3}.

\end{thebibliography}

\newpage

\appendix
\section*{Appendix}

\section{Satellite Images as Features}
We downloaded the following satellite imagery products for the year 2018 via the Google Earth Engine \citep{gorelick2017google} platform.

\begin{itemize}
\item \textbf{Vegetation.}The Enhanced Vegetation Index measures thickness of vegetation in areas, with the index particularly modified from the NDVI formula for improved sensitivity, which is available from NASA MODIS at a resolution of 1km
\item \textbf{Aridity.} The Global Aridity Index \citep{trabucco_zomer_2019} measures evapotranspiration and rainfall deficit, and is available from the Consortium of Spatial Information at a resolution of 30 arc-seconds (close to 1km per tile).
\item \textbf{Temperature.} Land Surface Temperature measures on-ground temperature during the day, and is available from NASA MODIS at a resolution of 1km.
\item \textbf{Nighttime Lights.} The Nighttime Day/Night Band Composites of the Visible Infrared Imaging Radiometer Suite (VIIRS DNB) measures radiance level at night, and is available from the National Oceanic and Atmospheric Administration at a resolution of 500m;
\item \textbf{Population.} The WorldPop Grid Square Population Data disaggregates to grid squares census-based population counts, and is available from the WorldPop Global project at a resolution of 100m;
\item \textbf{Elevation.} Digital Elevation measures elevation at a global scale, and is available from the Shuttle Radar Topography Mission (SRTM) at a resolution of 30m;
\item \textbf{Urban Change.} Change Year Index measures how recently the area has been converted to impervious (e.g. development of a new settlement), and is available from TsingHua University at a resolution of 30m.

\end{itemize}

\section{Relationship of Covariates to WASH access}
\begin{figure*}[h!]
  \centering
  \includegraphics[width=0.61\linewidth]{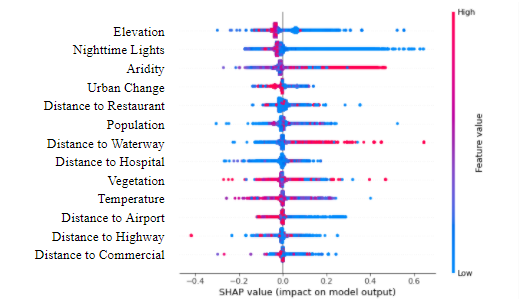}
\caption{Feature impact on percent households with no water supply.}
\label{fig:shap_water}
\end{figure*}

\begin{figure*}[h!]
  \centering
  \includegraphics[width=0.61\linewidth]{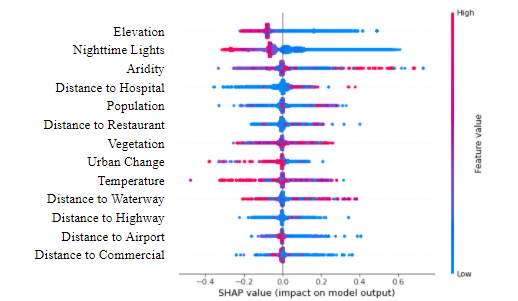}
\caption{Feature impact on percent households with no connection to sewage.}
\label{fig:shap_sewage}
\end{figure*}

We explain the relationship between predictors and the predicted variables using SHAP \citep{lundberg2017unified}, a method that uses game theory to measure the marginal contribution of each predictor variable to the resulting prediction. For all 3 predicted variables, we observe nighttime lights to have a positive relationship with access to WASH. This agrees with previous studies wherein nighttime lights were shown to be good proxies for economic development. For access to water (Figure \ref{fig:shap_water}), we observe distance to the nearest waterway to have a negative relationship with access, with farther distance associated with less access. This makes sense because households nearer bodies of water like rivers and lakes will most definitely have more access to water. For access to sewage (Figure \ref{fig:shap_sewage}), we observe the recency of urbanization (as measured by the urban index), with older urbanized areas associated with higher access to sewage. This affirms domain knowledge because the older the area has been since it was developed into an urban area, the more likely the area is developed with the proper infrastructure.
\newpage

 \begin{figure*}[h!]
  \centering
  \includegraphics[width=0.61\linewidth]{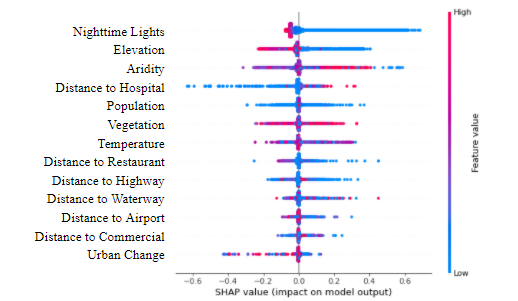}
\caption{Feature impact on percent households with no toilet.}
\label{fig:shap_toilet}
\end{figure*}

\newpage

\end{document}